\documentclass[twocolumn,twoside,preprintnumbers,superscriptaddress,nofootinbib,natbib,bm,amsrefs]{revtex4-2}

\usepackage{multirow,array}
\usepackage{amsmath,amssymb}
\usepackage{graphicx}
\usepackage{color}
\usepackage{slashed}
\usepackage{footmisc}
\usepackage{comment}
\usepackage{braket}
\usepackage{colortbl}
\usepackage{orcidlink}
\usepackage{physics}
\usepackage{url}
\usepackage{placeins}

\definecolor{nicered}{rgb}{0.5,0.,0.}
\definecolor{nicegreen}{rgb}{0.,0.5,0.}
\definecolor{niceblue}{rgb}{0.,0.,0.5}

\usepackage{hyperref}

\hypersetup{colorlinks=true,citecolor=nicegreen,linkcolor=nicered,urlcolor=niceblue}

\allowdisplaybreaks

\begin{document}

\title{Feynman Integrals Meet Second-Order Partial Differential Equations}

\author{Xiang Chen} 
\affiliation{
Physik-Institut, Universit\"{a}t Z\"{u}rich, 8057 Z\"{u}rich, Switzerland
}

\author{Hantian Zhang} 
\email{hantian.zhang@cern.ch}
\affiliation{
Theoretical Physics Department, CERN, 1211 Geneva 23, Switzerland
}

\preprint{CERN-TH-2026-170, ZU-TH 25/26}

\begin{abstract}
We propose a second-order partial differential equation method to solve multi-loop Feynman integrals as an equilibrium problem.
As a proof-of-concept demonstration, we perform a Galerkin discretization of the corresponding variational form and employ the finite element method to compute two-loop four-point Feynman integrals.
This method can solve the integral over a broad region of multi-dimensional phase space once and for all. 
This work establishes a new connection between perturbative quantum field theory and modern partial differential equation methods, with the potential to streamline a wide range of phenomenological applications.
\end{abstract}

\maketitle

\section*{Introduction}
The linear system of first-order differential equations~(DEs) has become the method of choice for computing multi-loop Feynman integrals~\cite{Kotikov:1990kg,Remiddi:1997ny,Gehrmann:1999as} in a wide range of challenging phenomenological applications in high-energy physics.
In the past decade, this method has been significantly advanced in various directions. For example, the $\epsilon$-factorized canonical form~\cite{Henn:2013pwa} has greatly facilitated the analytic calculation of multi-loop and high-multiplicity QCD corrections~\cite{Caola:2020dfu,Bargiela:2021wuy,Gehrmann:2024tds,Chicherin:2018old,Abreu:2018zmy,Badger:2019djh,Abreu:2023rco,Liu:2024ont,Chicherin:2025mvc,Henn:2024ngj,Abreu:2024fei,Henn:2025xrc,Liu:2026hdp}, 
while the generalized series expansion~\cite{Lee:2017qql,Moriello:2019yhu} with boundary conditions determined in asymptotic limits has led to the auxiliary mass flow method~\cite{Liu:2017jxz}, the semi-analytic method~\cite{Fael:2021kyg,Fael:2022rgm,Armadillo:2022ugh} and the analytic expansion method~\cite{Davies:2018qvx,Davies:2022ram}, which have been successfully applied to two-loop four-point electroweak corrections~\cite{Chen:2022mre,Armadillo:2022bgm,Bi:2023bnq,Davies:2026wbx}.

Looking ahead, the high-precision experimental measurements at future $e^+e^-$ colliders, such as \mbox{FCC-ee}~\cite{FCC:2018byv,FCC:2018evy,FCC:2025lpp} and CEPC~\cite{CEPCStudyGroup:2018rmc,CEPCStudyGroup:2018ghi}, require unprecedented precision of theoretical predictions, where higher-order electroweak and QCD corrections to high-multiplicity processes are needed for differential observables.
These differential calculations are essential for correctly interpreting the experimental data~\cite{deBlas:2025gyz}, for example, in extracting electroweak precision observables~\cite{Denner:2005fg}, studying QCD phenomenology~\cite{Chen:2026jxf}, and precision new physics searches~\cite{Bredt:2025zxc}.
However, the theoretical precision required for future $e^+e^-$ colliders is beyond the reach of current techniques. In particular, differential predictions require evaluating high-multiplicity amplitudes over broad regions of a vast multi-dimensional phase space rather than at isolated kinematical points. This challenge is further compounded by the fact that the experiments probe directly at the electroweak scale, where the top quark, Higgs boson, $W$, $Z$ and Goldstone bosons all appear in multi-loop diagrams.
Therefore, new techniques must be developed to meet these challenges, and  the first steps have been taken by exploring an efficient way of computing the series expansion through spectral methods with Chebyshev polynomials~\cite{Liu:2026cpf,Abreu:2026vxw} in the canonical form.
This is a promising direction that matches the nature of first-order DEs as a transport problem along one-dimensional paths.

In this paper, we propose a global approach for computing Feynman integrals over the multi-dimensional phase space by formulating a variational form of the linear system of second-order partial differential equations~(PDEs) in an arbitrary basis.
Upon Galerkin discretization, the problem is reduced to solving a linear system of equations over a bounded phase space in a single computation.
This procedure translates the Feynman integral calculations into an equilibrium problem, which is different from the path-based transport methods~\cite{Lee:2017qql,Moriello:2019yhu,Liu:2017jxz,Fael:2021kyg,Fael:2022rgm,Armadillo:2022ugh,Hidding:2020ytt,Liu:2022chg,Prisco:2025wqs,PetitRosas:2025xhm,Baur:2026zlw,Czakon:2026tog,Liu:2026cpf,Abreu:2026vxw,Huang:2026rjb}.
Numerical methods for solving the second-order PDEs are an active field of research in mathematics and continue to advance across a wide range of applications in the natural sciences and engineering.
As a proof-of-concept demonstration, we employ the finite element method~(FEM)~\cite{FEMbook} to compute the two-loop four-point Feynman integrals, which possess a two-dimensional kinematical phase space.

\section*{Methodology}
We focus on the four-point Feynman integrals with two Mandelstam variables $s$ and $t$, in order to keep the following discussion structured.
The generalization to high-multiplicity cases is conceptually simple but requires substantial technical efforts.
For a vector of master integrals~(MIs) $\vec{I}$, we construct the second-order PDEs 
\begin{align}
    \nabla^2  \vec{I} := (\partial^2_s + \partial^2_t)\, \vec{I} = M(s,t, \epsilon) \, \vec{I} \,,
    \label{eq:2ndPDE}
\end{align}
where $\nabla = (\partial_s,\partial_t)$ is the gradient operator, $\epsilon$ is  the dimensional regulator, and $M$ is the matrix for the linear system.
The naturalness of the Laplacian operator $\nabla^2$ is discussed in Appendix~\ref{app:laplacian}.
Note that since we can take derivatives and apply integration-by-parts~(IBP) reductions iteratively, the IBP relations required for constructing the second-order PDEs are identical to the ones for first-order DEs. Therefore, no additional computational cost occurs in the reduction step.
For a particular uncoupled MI $I_n$ ordered by the sectors, we have
\begin{align}
    \nabla^2 I_n =  V(s,t,\epsilon) \, I_n + f(s,t,\epsilon,\{ I_1, \dots, I_{n-1} \}) \,,
    \label{eq:pdeform}
\end{align}
where we denote $V$ as the potential function and $f$ as the load function, which depends on lower-sector MIs that are pre-computed before $I_n$.
As mentioned before, this is an equilibrium problem that can be solved by the Galerkin discretization on its variational form over a bounded two-dimensional mesh $\mathcal{M}$ with domain $\Omega$ and boundary $\partial\Omega$.

The variational formulation requires the introduction of the so-called test function $v$ that vanishes on the boundary $v_{|\partial\Omega} = 0$, which assumes to be square integrable in its first derivative.
The variational form can be obtained by integrating over the domain and applying IBP such that
\begin{align}
& -\int_{\Omega} \nabla  I_n \cdot \nabla v - \int_{\Omega} V \, I_n \, v = \int_{\Omega} f \, v \,, \\[2mm]
& \qquad I_n|_{\partial \Omega} = \gamma_n(s,t,\epsilon) \;\; \mbox{and} \;\; v|_{\partial \Omega} = 0 \,.
    \label{eq:weakform}
\end{align}
where boundary conditions $\gamma_n(s,t,\epsilon)$  are input data.
For various cases where analytic boundary conditions are not available, they can be computed numerically to high precision using path-based transport methods~\cite{Lee:2017qql,Moriello:2019yhu,Liu:2017jxz,Fael:2021kyg,Fael:2022rgm,Armadillo:2022ugh,Hidding:2020ytt,Liu:2022chg,Prisco:2025wqs,PetitRosas:2025xhm,Baur:2026zlw,Czakon:2026tog,Liu:2026cpf,Abreu:2026vxw,Huang:2026rjb}. 
Usually, we  compute $I_n$ and  $\gamma_n$ in an $\epsilon$-expanded Laurent series
\begin{align}
    I_n = \sum_{i} \epsilon^i \, I_{n}^{(i)} \,, \quad 
    \gamma_n = \sum_{i} \epsilon^i \, \gamma_{n}^{(i)}\,,
\end{align}
to certain order of $\epsilon$ for phenomenological applications.
In order to compute $I_n$ with non-homogeneous Dirichlet boundary conditions, we can decompose it into
\begin{align}
    &I_n = u + u_0 \;\; \text{with} \;\; u_0|_{\partial \Omega} = \gamma_n, \, u|_{\partial\Omega} = u_0|_{\Omega/\partial\Omega} = 0 \, ,
    \label{eq:funcdecomp}
\end{align}
where $u_0$ is the offset function with support near the boundary, while $u$ represents the function over the whole domain with vanishing boundary conditions.
In the variational formulation, both functions $u$ and $v$ belong to the Sobolev space
\begin{align}
    H^{1}_{0}(\Omega) := \Big\{u: \Omega \to \mathbb{C} \; \Big| \; u|_{\partial\Omega} = 0\, , \, \int_{\Omega} |\nabla \,u|^2 < \infty \Big\},
    \label{eq:sobolev}
\end{align}
thereby relaxing the regularity requirements on the solution compared to the original PDE formulation in Eq.~\eqref{eq:pdeform}.

In the next step, we perform the Galerkin discretization and employ the linear FEM for simplicity.
We discretize the mesh $\mathcal{M}$ into triangle elements $\mathcal{K}$ in the domain and line elements on the boundary.
The mesh $\mathcal{M}$ contains $N$-ordered nodes $\mathcal{N(\mathcal{M})} = \{ p_1, \dots,p_N \}$ where $p_i$ represents the coordinate in the $(s,t)$-plane.
We denote the number of boundary nodes on $\partial\Omega$ by $N_{\rm bdy}$ and of interior nodes on $\Omega/\partial\Omega$ by $N_{\rm int}$.
We assign the nodal basis $\mathcal{B}_N = \{b_N^i\}$  as the support for functions over the whole domain
\begin{align}
    & u = \sum_i \mu_i b^i_N \,, \; v = \sum_j \mu_j b^j_N \,, \;\; \mbox{with} \;\; b_N^i(p_k) = \delta^{i}_{k}\,,
    \label{eq:nodal}
\end{align}
where the coefficient vector $\vec{\mu} = \{\mu_i\}$ is the object to be solved.
Note that the nodal basis also supports the offset function $u_0$ on the boundary, but their coefficients are directly determined by the boundary conditions.
For each local triangle element $\mathcal{K}$, the nodal basis can be parametrized by barycentric functions $(\lambda_1, \lambda_2, \lambda_3)$ such that the relation~\eqref{eq:nodal} is satisfied and the center of mass of the triangle is located at $(\lambda_1, \lambda_2, \lambda_3) = (\frac{1}{3},\frac{1}{3},\frac{1}{3})$.
In the linear FEM, the nodal basis is approximated by degree-one polynomials constructed from the barycentric functions, providing a piecewise-linear approximation of the function on the mesh.
The component of the symmetric Galerkin matrices for the bilinear terms, including  derivative and potential functions, and of the vector for the load function, can be computed as
\begin{align}
    \label{eq:GlobalMatrix}
   \mathbf{A}^{ij} & = -\sum_{\mathcal{K}}\int_{\Omega_{\mathcal{K}}} \nabla b_N^i   \cdot \nabla b_N^j \,,\\
   \mathbf{B}^{ij} & = -\sum_{\mathcal{K}}\int_{\Omega_{\mathcal{K}}}  V \, b_N^i \,  b_N^j \,,\\
   \mathbf{F}^{i} & =   \sum_{\mathcal{K}}\int_{\Omega_{\mathcal{K}}} f \, b_N^i\,.
   \label{eq:GlobalVector}
\end{align}
The integration over local triangle elements can be performed analytically for the derivative term and approximated by Gaussian quadratures for the potential and load functions.
The global Galerkin matrix can be assembled in an element-oriented way from the local computations on the elements.
We can assemble the FEM system of equations
for the uncoupled PDE in Eq.~\eqref{eq:weakform} as
\begin{align}
    \begin{pmatrix}
     \mathbf{I} & \mathbf{0} \\
     \mathbf{A}_\partial + \mathbf{B}_\partial  & \mathbf{A}_{\rm int} + \mathbf{B}_{\rm int} \\
    \end{pmatrix}
    \begin{pmatrix}
        \vec{\mu}_\partial \\
        \vec{\mu}_{\rm int}
    \end{pmatrix} 
    & = 
    \begin{pmatrix}
        \vec{\gamma}_n \\
        \vec{F}
    \end{pmatrix}\,,
    \label{eq:FEM}
\end{align}
where $\mathbf{A}_{\rm int},\mathbf{B}_{\rm int}$ are $N_{\rm int} \times N_{\rm int}$ matrices and $\mathbf{A}_{\partial},\mathbf{B}_{\partial}$ are $N_{\rm int} \times N_{\rm bdy}$ matrices, which are both subsets of the global matrices $\mathbf{A}$ and $\mathbf{B}$. 
In this notation, we have $\vec{\mu}=\{\vec{\mu}_\partial,\vec{\mu}_{\rm int}\}$.
We note that $\mathbf{A}_{\rm int},\mathbf{B}_{\rm int}$ and $\mathbf{A}_{\partial},\mathbf{B}_{\partial}$ are all sparse matrices, and the FEM system can be efficiently solved using sparse linear solvers.

For the coupled system, we discuss a simple scenario with two coupled PDEs of $I_{c1}$ and $I_{c2}$. The variational form for the coupled system is
\begin{align}
& -\int_{\Omega} \nabla  I_{c1} \cdot \nabla v - \int_{\Omega} (V_{c1}^{(c1)} \, I_{c1} + V_{c1}^{(c2)} \, I_{c2} ) \, v = \int_{\Omega} f_{c1} \, v \,, \nonumber \\
& -\int_{\Omega} \nabla  I_{c2} \cdot \nabla v - \int_{\Omega} (V_{c2}^{(c1)} \, I_{c1} + V_{c2}^{(c2)} \, I_{c2} ) \, v = \int_{\Omega} f_{c2} \, v \,. 
    \label{eq:coupleweakform}
\end{align}
Its discretized form is
\begin{widetext}
\begin{align}
    \begin{pmatrix}
     \mathbf{I} & \mathbf{0}  & \mathbf{0} & \mathbf{0} \\[2pt]
     \mathbf{A}_{c1,\partial} +  \mathbf{B}_{c1,\partial}^{(c1)} &
     \mathbf{A}_{c1,\rm int} +  \mathbf{B}_{c1,\rm int}^{(c1)} & \mathbf{B}_{c1,\partial}^{(c2)} & \mathbf{B}_{c1, \rm int}^{(c2)} \\[2pt]
    \mathbf{0}  & \mathbf{0} & \mathbf{I} & \mathbf{0} \\[2pt]
    \mathbf{B}_{c2,\partial}^{(c1)} & \mathbf{B}_{c2,\rm int}^{(c1)} & 
    \mathbf{A}_{c2,\partial} +  \mathbf{B}_{c2,\partial}^{(c2)} & \mathbf{A}_{c2,\rm int} +  \mathbf{B}_{c2,\rm int}^{(c2)} &
    \end{pmatrix}
    \begin{pmatrix}
        \vec{\mu}_{c1,\partial} \\[4pt]
        \vec{\mu}_{c1} \\[4pt]
        \vec{\mu}_{c2,\partial} \\[4pt]
        \vec{\mu}_{c2}
    \end{pmatrix} 
    & = 
    \begin{pmatrix}
        \vec{\gamma}_{c1} \\[4pt]
        \vec{F}_{c1} \\[4pt]
        \vec{\gamma}_{c2} \\[4pt]
        \vec{F}_{c2}
    \end{pmatrix}\,,
    \label{eq:FEMcoupled}
\end{align}
\end{widetext}
where the coupled effects are represented by the off-diagonal block matrices $\mathbf{B}_{c1, \partial}^{(c2)}, \mathbf{B}_{c2, \partial}^{(c1)},\mathbf{B}_{c1, \rm int}^{(c2)}, \mathbf{B}_{c2, \rm int}^{(c1)}$, which are computed from the potential functions $V_{c1}^{(c2)}, V_{c2}^{(c1)}$.
The extension to a more complicated coupled system of PDEs is straightforward.

\section*{Numerical Experiments}
\begin{figure}[b]
    \centering
    \includegraphics[width=0.4\linewidth]{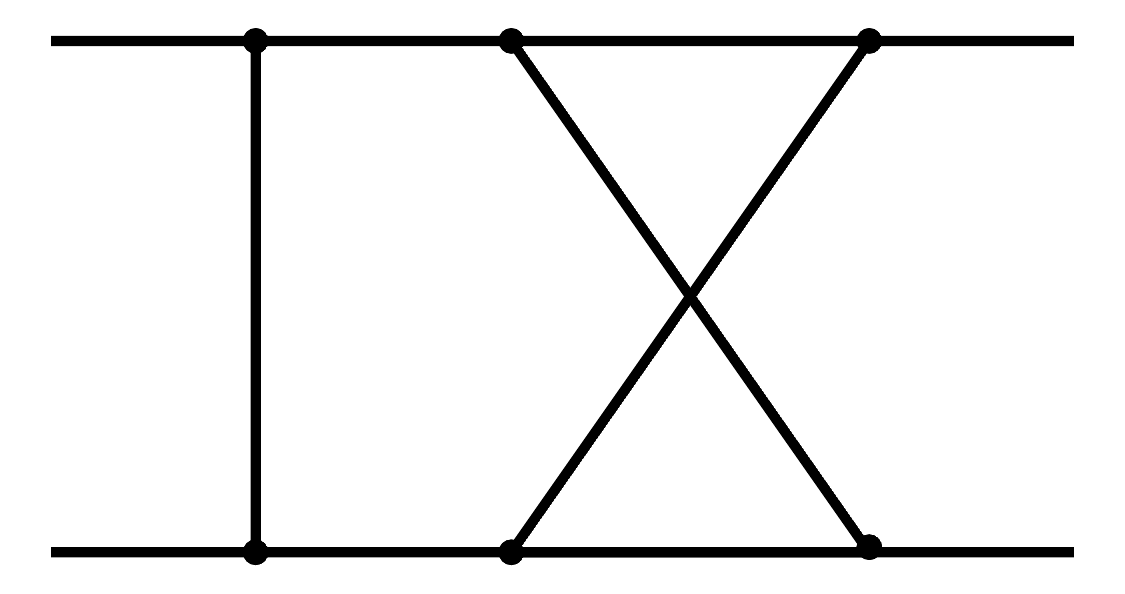}
    \caption{Feynman diagram of massless non-planar integral.}
    \label{fig:feyndiag}
\end{figure}

In the following, we present the numerical experiments on a massless two-loop four-point non-planar Feynman integral family shown in Fig.~\ref{fig:feyndiag}.
This integral was first computed analytically to $\mathcal{O}(\epsilon^0)$ in Ref.~\cite{Tausk:1999vh} and can also be obtained using the canonical form approach.
The purpose of these numerical experiments is not to apply the FEM approach to cutting-edge computations where analytic results are unavailable, 
but to use an analytically well-controlled two-loop example to illustrate the global approach over a broad region of phase space based on second-order PDEs and benchmark its numerical performance.
We use \textsc{LiteRed}~\cite{Lee:2013mka} to perform the IBP reduction to a non-canonical basis with 12 MIs, where the top sector contains 2 coupled MIs.
We denote the corner integral of the top-sector shown in Fig.~\ref{fig:feyndiag} by $I_{\rm NPL}$, which couples to another top-sector integral with an irreducible numerator, denoted by $I_{\rm NPL}^{(\rm num)}$.
The analytic solutions of these 12 MIs are taken from \textsc{AsyInt}~\cite{Zhang:2024fcu} and serve as the boundary conditions and as benchmarks for comparison with the numerical results on the interior nodes.
In order to obtain $\mathcal{O}(\epsilon^0)$ terms of coupled MIs, we perform $\epsilon$-expansion on each MI which yields in total 55 PDEs with non-zero load functions, all of which are solved using the FEM. 
We implement the FEM in \textsc{Mathematica} and use the  \texttt{LinearSolve} function to solve the sparse system of equations.

\begin{table}[b]
    \centering 
    \renewcommand{\arraystretch}{1.35}
    \begin{tabular}{|c|c|c|c|c|} \hline
     \multicolumn{5}{|c|}{FEM Experiment No.~1 
     } \\ \cline{1-5}
        ~$N_{\rm bdy}$~ & ~$N_{\rm int}$~ & $\varepsilon_{\rm avg}$   & $\varepsilon_{\rm max}$ &  Runtime [sec]  \\ \hline
        158 & 904 &   $1.79\times 10^{-3}$ & $2.60\times 10^{-2}$ &  0.14   \\ 
        \rowcolor{gray!15}
        177 & 1885  & $1.00\times 10^{-3}$ & $1.95\times 10^{-2}$ & 0.28 \\ 
         338 & 4769 & $4.65\times 10^{-4}$ & $1.21\times 10^{-2}$ & 0.75 \\ 
        \rowcolor{gray!15}
        376 & 9666 & $2.70\times10^{-4}$ & $1.02\times 10^{-2}$ & 1.56 \\ 
        631 & 16201 & $1.92\times 10^{-4}$ & $8.53\times 10^{-3}$ & 2.75 \\ \hline
    \end{tabular}
    \caption{First FEM experiment on uniform meshes with phase space cuts $p_T > 50$~GeV and $\sqrt{s}< 200$~GeV. The average and maximal errors $\varepsilon_{\rm avg}$ and $\varepsilon_{\rm max}$ for $\mathcal{O}(\epsilon^0)$ term of $I_{\rm NPL}$ compared with the analytic results are reported.
    The runtimes of \texttt{LinearSolve} for solving the coupled system of $I_{\rm NPL}$ and $I_{\rm NPL}^{(\rm num)}$ from $\mathcal{O}(\epsilon^{-4})$ to $\mathcal{O}(\epsilon^{0})$ are reported.
    The benchmarks are performed on a single core of \textsc{Apple M4} CPU using \textsc{Mathematica 14.3} with double-precision arithmetic.
    }
    \label{tab:exp1}
\end{table}

In the first numerical experiment, we establish the benchmarks on uniform meshes, where all triangle elements have similar measures, over a phase space region with $p_T > 50$~GeV and $\sqrt{s}< 200$~GeV cuts. 
This integral exhibits strong singularities in the $p_T \to 0$, $\sqrt{s} \to 0$ limit, leading to hierarchical numerical values of several orders of magnitude.
Therefore, it is advantageous to first benchmark the performance on a small phase space region. 
To quantify the numerical performance, we define the average and maximal errors by comparing the FEM results with the exact analytic results,
$
    \varepsilon_{\rm avg}  =\frac{1}{N_{\rm int}}\sum_{i}^{N_{\rm int}} \left|\frac{I^{(\epsilon^0)}_{\rm exact}(s_i,t_i) - I^{(\epsilon^0)}_{\rm FEM}(s_i,t_i)}{I^{(\epsilon^0)}_{\rm exact}(s_i,t_i)}\right| $ and
$
    \varepsilon_{\rm max}  = \max \left|\frac{I^{(\epsilon^0)}_{\rm exact}(s_i,t_i) - I^{(\epsilon^0)}_{\rm FEM}(s_i,t_i)}{I^{(\epsilon^0)}_{\rm exact}(s_i,t_i)}\right| \,,
$
where the absolute value is equivalent to the $L^2$-norm for complex numbers.
The runtimes of the \texttt{LinearSolve} function for solving 
the coupled system of PDEs involving the $\mathcal{O}(\epsilon^{-4})$ to $\mathcal{O}(\epsilon^{0})$ terms of $I_{\rm NPL}$ and $I_{\rm NPL}^{(\rm num)}$ are profiled.
Note that the system at $\mathcal{O}(\epsilon^n)$ is not coupled to the other orders but depends on the solutions from $ \mathcal{O}(\epsilon^{-4})$ to $\mathcal{O}(\epsilon^{n-1})$, whereas within $\mathcal{O}(\epsilon^n)$ the system is coupled between $I_{\rm NPL}$ and $I_{\rm NPL}^{(\rm num)}$.
The corresponding numerical 
results for $I_{\rm NPL}$  are summarized in Table~\ref{tab:exp1}. Increasing the
number of interior nodes from $904$ to $16201$ reduces the average error
from $1.79\times10^{-3}$ to $1.92\times10^{-4}$, while the maximal
error decreases from $2.60\times10^{-2}$ to $8.53\times10^{-3}$.
Both the average and maximal errors exhibit an algebraic convergence pattern
\begin{align}
    \varepsilon \sim \mathcal{O}(N_{\rm  int}^{-\alpha})\,,
\end{align}
where the convergence rate $\alpha$ is estimated by linear regression on the logarithmic scale.
For the first experiment, they are estimated to 
$ \alpha^{(1)}_{\rm avg} = 0.78 $ and $ \alpha^{(1)}_{\rm max} = 0.39$.
The numerical results for $I_{\rm NPL}^{(\rm num)}$ are similar, with average and maximal errors agreeing with those of $I_{\rm NPL}$ within a few percent. 
Since the two integrals are solved simultaneously as a coupled system, similar numerical accuracies are expected, and we therefore do not present the results separately.
Over the mesh refinement, 
the boundary-to-interior ratio $N_{\rm bdy}/N_{\rm int}$ decreases from 17.5\% to 3.9\% and 
the runtime increases
from $0.14$ to $2.75$ seconds, demonstrating the computational efficiency of the global FEM approach.
The runtime scales approximately linearly with the total number of mesh nodes,
indicating the sparsity of the Galerkin matrices.
We also conduct numerical tests on graded meshes, where the measure of triangle elements increases with $\sqrt{s}$.
Over a similar mesh-refinement range, the tests on graded meshes yield results comparable to those on uniform meshes, with average and maximal errors larger by about 10\% and 2\%, while maintaining similar convergence rates and runtimes.

\begin{figure}[t]
    \centering
    \includegraphics[width=0.95\linewidth]{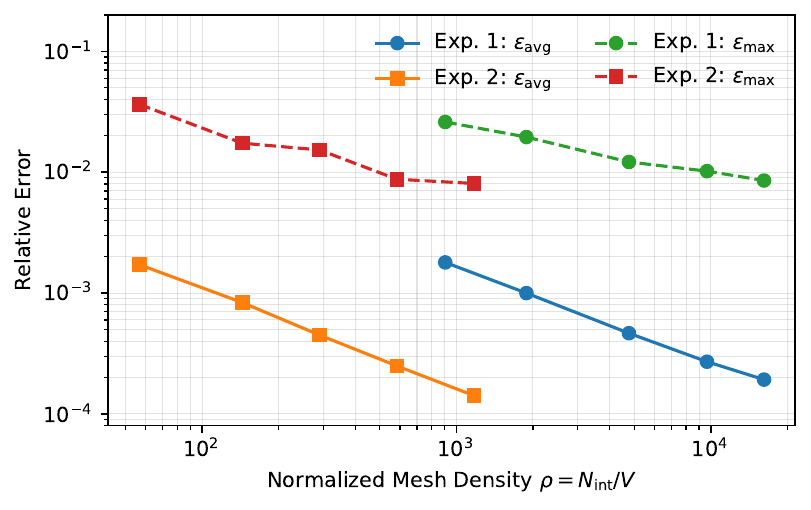}
    \caption{Convergence in terms of average and maximal errors for $I_{\rm NPL}$ at $\mathcal{O}(\epsilon^0)$ on normalized mesh density for the first and the second FEM experiments.}
    \label{fig:femconvergence}
\end{figure}

\begin{figure}[t]
    \centering
    \includegraphics[width=0.9\linewidth]{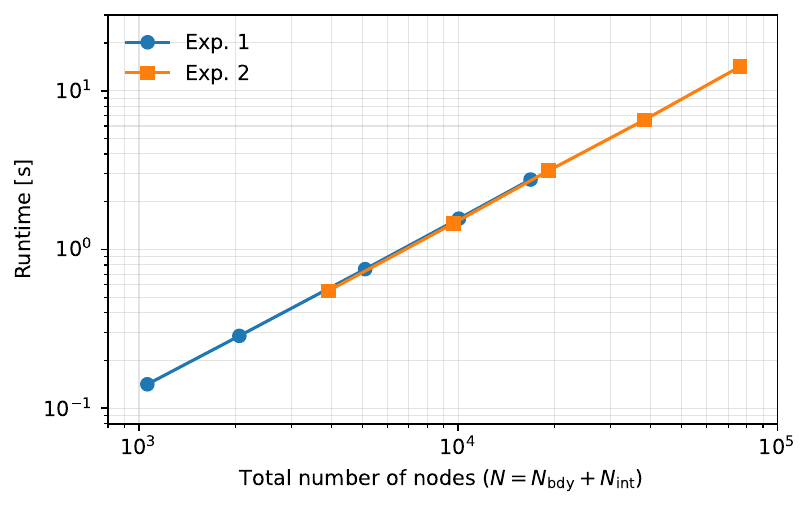}
    \caption{Runtime of \texttt{LinearSolve} for solving the coupled system of PDEs involving $I_{\rm NPL}$ and $I_{\rm NPL}^{(\rm num)}$ from $\mathcal{O}(\epsilon^{-4}) $ to $ \mathcal{O}(\epsilon^{0}) $.
    They are computed on a single core of \textsc{Apple M4} CPU using \textsc{Mathematica 14.3} with double-precision arithmetic.
    }
    \label{fig:femruntime}
\end{figure}

\begin{table}[t]
    \centering 
    \renewcommand{\arraystretch}{1.35}
    \begin{tabular}{|c|c|c|c|c|} \hline
     \multicolumn{5}{|c|}{FEM Experiment No.~2
     } \\ \cline{1-5}
        ~$N_{\rm bdy}$~ & ~$N_{\rm int}$~ & $\varepsilon_{\rm avg}$   & $\varepsilon_{\rm max}$ &  Runtime [sec]  \\ \hline
         290 & 3627 &   $1.71\times 10^{-3}$ & $3.64\times 10^{-2}$ &  0.54   \\ 
        \rowcolor{gray!15}
         441 &  9217 & $8.34\times 10^{-4}$ & $1.74\times 10^{-2}$ &  1.46 \\ 
         670 & 18518 & $4.50\times 10^{-4}$ & $1.53\times 10^{-2}$ &  3.14 \\ 
          \rowcolor{gray!15}
         910 & 37386 & $2.49\times 10^{-4}$ & $8.75\times 10^{-3}$ & 6.55  \\ 
         1328 & 75008 & $1.41\times 10^{-4}$ & $8.06\times 10^{-3}$ & 14.20  \\ \hline
    \end{tabular}
    \caption{Second FEM experiment on graded meshes with phase space cuts $p_T > 100$~GeV and $\sqrt{s}< 1000$~GeV.
    The average and maximal errors for $I_{\rm NPL}$ and runtimes for solving the coupled $I_{\rm NPL}$-and-$I_{\rm NPL}^{(\rm num)}$ system in \textsc{Mathematica} are reported. The setup is the same as described in Table~\ref{tab:exp1}.
    }
    \label{tab:exp2}
\end{table}
In the second numerical experiment, 
we consider a substantially larger phase-space region with
$p_T>100~\mathrm{GeV}$ and $\sqrt{s}<1000~\mathrm{GeV}$ using graded
meshes. 
This phase space region is 64 times larger than that considered in the first experiment, and the solution is less singular.
As reported in Table~\ref{tab:exp2}, increasing
the number of interior nodes from $3627$ to $75008$ reduces the average
error from $1.71\times10^{-3}$ to $1.41\times 10^{-4}$ and the maximal
error from $3.64\times10^{-2}$ to $8.06\times 10^{-3}$,
yielding the algebraic convergence rates  
$\alpha^{(2)}_{\rm avg} = 0.83$ and $\alpha^{(2)}_{\rm max} = 0.50$.
The runtime again scales approximately linearly with the total number of mesh nodes, while the boundary-to-interior ratio decreases to $1.8\%$.
The convergence of both experiments is compared in
Fig.~\ref{fig:femconvergence} as a function of the normalized mesh
density $\rho=N_{\rm int}/V$, with $V=1$ and $64$ for the
first and second experiments, respectively. 
Due to a better regularity of the solution in the second experiment, it achieves the same level of accuracy as the first experiment with a lower mesh density.
Figure~\ref{fig:femruntime} shows that the runtimes of the two
experiments scale approximately linearly with the total number of mesh nodes
\begin{align}
    {\rm runtime} \sim \mathcal{O}\big(N^{1.08}\big)\,.
\end{align}
This scaling is consistent in both experiments, demonstrating the robustness and computational efficiency of the FEM approach over different phase space regions.

Finally, we comment on the second-order PDE approach.
The current linear FEM with degree-one polynomial nodal basis functions is the simplest member of the FEM family, yielding robust and convergent results.
This method can be systematically improved in various directions. For example, the higher-degree polynomial approximation and the incorporation of the spectral method for the nodal basis, known as the $hp$-FEM~\cite{GuoBabuska1986a} and the spectral element method~\cite{Patera1984}, can greatly improve the algebraic convergence rate.
The latter has the potential to reach an exponential convergence rate over phase space regions where the Feynman integrals are regular.
Besides the regularity of the solution, the convergence rate of the FEM is further influenced by the choice of MI basis and the mesh design and refinement.
While the method is applicable to arbitrary choices of MIs, we observe that selecting MIs with irreducible numerators yields better numerical accuracy than using MIs with dotted propagators.
%

For Feynman integrals involving massive internal propagators,
the threshold behavior arising from cutting on-shell propagators with real-valued masses can also be accommodated by the FEM approach.
Although the PDE becomes singular at the cut threshold, the integral itself remains continuous, exhibiting only a kink, and still belongs to the Sobolev space required by the variational formulation.
However, this reduced regularity deteriorates the numerical convergence.
This numerical issue can be addressed either by partitioning the phase space along the threshold or by introducing the complex-mass scheme~\cite{Denner:2005fg,Actis:2016mpe,Buccioni:2019sur,Armadillo:2022ugh,Armadillo:2026wdb} to regularize the cut-threshold behavior.
We have tested both approaches with real- and complex-valued masses on a massive one-loop four-point integral and find excellent convergence rates.
%

\section*{Summary}
In summary, we have introduced a novel approach to solve Feynman integrals by formulating a linear system of second-order PDEs in its variational form.
This formulation turns the Feynman integral problem into an equilibrium problem, which can be efficiently solved using a mesh-based Galerkin discretization.
This is a global approach that solves the Feynman integrals over a region of multi-dimensional phase space in a single computation, whereas the traditional first-order DE approach relies on the one-dimensional path-based transport through series expansions.
As a proof-of-concept demonstration, we have employed the linear FEM to solve the two-loop four-point Feynman integrals over different regions of the phase space, requiring only a limited number of boundary conditions as input.
This work establishes a new connection between perturbative quantum field theory and modern PDE methods. It provides a promising direction for addressing the theoretical challenges posed by the high-precision measurements at future $e^+e^-$ colliders.

\vskip 0.6cm
\noindent\textbf{Acknowledgment:}
We are grateful to Samuel Abreu for valuable discussions and continuous support.
We also thank Afonso Guerreiro for discussions.
H.~Z. is funded by the European Union under the Marie Sk{\l}odowska-Curie Actions (MSCA) grant 101202083 -- ``HINOVA''.
X.~C. is supported by the Swiss National Science Foundation (SNSF) under contract 200020$\_$219367 and partly by the UZH Postdoc Grant No.~[FK-25-104].

\newpage
\appendix
\renewcommand{\theequation}{A.\arabic{equation}}
\newcommand{\MI}{\vec I}
\newcommand{\res}{\vec r}

\section{Naturalness of Laplacian Structure}
\label{app:laplacian}
The second-order Laplacian PDEs in Eq.~\eqref{eq:2ndPDE} are constructed directly, and their variational formulation is self-contained. 
As an independent perspective on the naturalness of this Laplacian structure, we show that the same PDEs are recovered at the minimum of a least-squares functional of the first-order DE residuals
\begin{equation}
 \mathcal J[\MI]=\frac12\sum_{i=s,t}\int_\Omega
 \res_i{}^\dagger \, \res_i\, ,
 \qquad
 \res_i=(\partial_i-A_i)\,\MI\, ,
 \label{eq:functional}
\end{equation}
where $A_i(s,t,\epsilon)$ are the first-order DE matrices and the exact MIs satisfy $\res_i=0$. Holding the Dirichlet boundary conditions, the variational principle gives
\begin{align}
 \delta\mathcal J
 &=\operatorname{Re}\sum_i\int_\Omega
 \left[(\partial_i-A_i)\delta\MI\,\right]^\dagger
 \res_i\, \notag\\
 &=-\operatorname{Re} \int_\Omega
  \delta\MI{\,}^\dagger  \sum_i(\partial_i+A_i^\dagger)
 \, \res_i\,,
 \label{eq:variation}
\end{align}
where the IBP relation is used in the second line.
Independent variations of the real and imaginary parts of MI functions yield the stationary condition
\begin{align}
 0&=-\sum_i(\partial_i+A_i^\dagger) \, \res_i \nonumber \\
  &=-\nabla^2\MI
   +\sum_i(A_i-A_i^\dagger)\, \partial_i\MI
   +\sum_i(\partial_iA_i+A_i^\dagger A_i)\, \MI \,.
 \label{eq:ls}
\end{align}
Using $\partial_i\MI=A_i\MI+\res_i$, 
the equation becomes
\begin{align}
 \nabla^2\MI = \underbrace{\sum_i(\partial_iA_i+A_i^2)}_{=\, M} \, \MI+\sum_i(A_i-A_i^\dagger)\, \res_i \,.
 \label{eq:decomposition}
\end{align}
The additional terms are proportional to the first-order residuals. For the exact MIs $\vec{r}_i =0$,  we obtain
\begin{equation}
 \nabla^2\MI=M\MI,
\end{equation}
as constructed in Eq.~\eqref{eq:2ndPDE}. Since $\mathcal J \geq 0$ and the exact MIs satisfying the physical Dirichlet conditions yield $\mathcal J=0$, the residual functional reaches its global minimum on these solutions. The Laplacian structure appearing in the directly constructed second-order PDEs is therefore independently recovered at the minimum of the least-squares functional of first-order DEs.

\bibliography{ref.bib}

\end{document}